\documentclass[12pt]{article}

\usepackage{paper2e}
\usepackage{mydefs2e}

\renewcommand{\d}{\partial}
\newcommand{\dd}{\raisebox{1.2pt}{$\stackrel{\raisebox{-1pt}%
{$\scriptscriptstyle\leftrightarrow$}}{\d}$}}
\newcommand{\N}{$\scr{N} = 1$\ }

\begin{document}

\begin{titlepage}
\preprint{UMD-PP-03-008}

\title{Five Dimensional Supergravity\\\medskip
in $\scr{N} = 1$ Superspace}

\author{W. D. Linch III,%
\footnote{{\tt ldw@physics.umd.edu}}%
\ \ Markus A. Luty,%
\footnote{{\tt mluty@physics.umd.edu}}%
\ \ J. Phillips%
\footnote{{\tt ferrigno@physics.umd.edu}}}

\address{Department of Physics, University of Maryland\\
College Park, Maryland 20742, USA}

\begin{abstract}
We give a formulation of linearized minimal 5-dimensional supergravity
in $\scr{N} = 1$ superspace.
Infinitesimal local 5D diffeomorphisms, local 5D Lorentz transformations,
and local 5D supersymmetry are all realized as off-shell superfield
transformations.
Compactification on an $S^1 / Z_2$ orbifold and
couplings to brane-localized supermultiplets are very simple in
this formalism.
We use this to show that 5-dimensional supergravity can naturally generate
$\mu$ and $B\mu$ terms of the correct size in gaugino- or radion-mediated
supersymmetry breaking.
We also include a self-contained review of linearized minimal 4D
supergravity in superspace.
\end{abstract}

\end{titlepage}

\section{Introduction}
Supergravity is naturally the messenger of supersymmetry
breaking in models where supersymmetry is broken in a hidden sector that
couples to the visible sector only through gravitational interactions
\cite{sugramed}.
Supergravity couples universally in the infrared, but theoretical
expectations (based on black holes and string theory) are that
the fundamental theory of gravity does not respect global symmetries.
In the low-energy effective theory, we therefore expect flavor-violating
contact terms between the hidden and observable sectors with gravitational
strength.
These couplings give flavor-dependent contributions to supersymmetry
breaking in the visible sector that are generally
the same size as the universal
contributions from gravity in the infrared,
leading to unacceptably large flavor-changing neutral currents.

This problem can be solved in `brane world' scenarios where the
visible and hidden sector are localized on spatially separated
3-branes \cite{hw,rs0}.
In these scenarios, the short-distance behavior of gravity does not
affect the transmission of supersymmetry breaking since it must
propagate over long distances \cite{rs0}.
Supersymmetry breaking is therefore dominated by infrared gravity,
which couples universally and hence can give flavor-independent
supersymmetry breaking masses, solving the supersymmetric flavor
problem.
This leads to a variety of models in which supergravity plays an
important role in supersymmetry breaking \cite{rs0,gravmed}.


The simplest brane-world models are 5-dimensional.
Models involving 5D supergravity have been analyzed in a
number of ways.
Work has been done using the on-shell formulation of 5D
supergravity \cite{onshell}, effective field theory techniques \cite{LS},
and the off-shell formulation of 5D supergravity \cite{gr} pioneered by
Zucker \cite{zucker} and further developed by a number of
authors \cite{offshell}. 
In this paper, we formulate 5D linearized supergravity completely
in terms of $\scr{N} = 1$ superfields.
This approach has recently been developed for global supersymmetry in
\Ref{nimasf} (see \cite{sfpioneer} for earlier work).
This means that the fields depend on 4D superspace coordinates
$(x^m, \th_\al)$ and a 5$^{\rm th}$ coordinate $x^5$.
In this formalism, the full 5D Lorentz invariance and supersymmetry
is not manifest, but the advantage is that coupling to 4D matter
localized on branes is simple.
Some partial results on 5D supergravity using $\scr{N} = 1$ superfields
have been obtained in \Ref{sf}.


In this paper, we give the complete linearized action for 5D supergravity
in terms of $\scr{N} = 1$ superfields.
Although global 5D Lorentz invariance and supersymmetry are not manifest,
the full infinitesimal 5D local Lorentz transformations, local 5D diffeomorphisms,
and local 5D supersymmetry transformations are
realized off-shell as superfield transformations.
The induced 4D supergravity multiplet on the branes is the standard
minimal $\scr{N} = 1$ multiplet, and so the couplings to brane-localized
matter is simple.
As an application of this formalism, we present an operator that gives
rise to realistic $\mu$ and $B\mu$ terms in the context of gaugino-mediated
supersymmetry breaking \cite{gMSB} or radion-mediated supersymmetry
breaking (third paper in \Ref{gravmed}).

This paper is organized as follows.
Section 2 contains a review
of 4D linearized supergravity in superspace,
including component results and invariant
couplings to matter fields.
Section 3 contains the main results of this paper.
We show how the 5D supergravity multiplet is
embedded in $\scr{N} = 1$ superfields, and give the superfield action.
We also show that the bosonic terms correctly reproduce 5D
linearized gravity along with kinetic terms for the 5D
graviphoton.
Section 4 applies these results to an $S^1/Z_2$ orbifold and 
section 5 gives our conclusions.

%
%

\section{4D Supergravity in Superspace}
In this section we review linearized 4D supergravity in superspace.
This formalism is due to Siegel and Gates \cite{sg},
and is reviewed in \Refs{review,buch}.%
\footnote{
We use the spinor conventions of \Ref{wb}.
We use bispinor notation
$\d_{\al\dot\al} = \si^m_{\al\dot\al} \d_m$, {\it etc\/}.}

\subsection{Gauge Transformations and Superfield Content}
The construction is closely analogous to the construction of \N supersymmetric
gauge theory in superspace.
We briefly review this construction for the case of a $U(1)$ gauge theory.
First, we demand that the gauge transformations take chiral superfields
into chiral superfields.
This restricts the gauge transformations to have the form
\beq
\de \Phi = i \Om \Phi.
\eeq
where $\Om$ is chiral.
Next, we note that antichiral superfields do not naturally transform under
this restricted gauge group, so that for example $\Phi^\dagger \Phi$ is
not gauge invariant.
We therefore introduce a gauge connection superfield $V$
and define covariant complex conjugation by
\beq
\Phi^\ddagger \equiv (1 + V + \scr{O}(V^2)) \Phi^\dagger.
\eeq
We want $\Phi^\ddagger$ to transform in the complex conjugate
representation with gauge parameter $\Om$:
\beq
\de \Phi^\ddagger = -i\Om \Phi^\ddagger.
\eeq
This requires that $V$ transforms as
\beq
\de V = -i(\Om - \Om^\dagger).
\eeq
We now know the superfield content and transformation laws, and can
work out the action and components.

We follow similar steps in the construction of linearized supergravity.
The starting point is the group of infinitesimal super-diffeomorphisms,
acting on a general superfield $\Psi$ as
\beq[superdiff]
\de \Psi = \La^\al D_\al \Psi + \La_{\dot\al} \bar{D}^{\dot\al} \Psi
+ \La^m \d_m \Psi.
\eeq
Here $\La_{\dot\al} \ne (\La_\al)^\dagger$ {\it a priori\/}.
The use of the differential operators $D_\al$ and $\bar{D}_{\dot\al}$ rather
than the ordinary derivatives $\d / \d \th^\al$ and $\d / \d \bar\th^{\dot\al}$
is not essential, but it makes it easy to keep global supersymmetry manifest.
The full group of super-diffeomorphisms \Eq{superdiff} is too large to give
a minimal formulation of supergravity.
We therefore restrict to the subgroup of diffeomorphisms that takes chiral
superfields to chiral superfields.
This gives the constraints
\beq[chiralconstraint]
\bar{D}_{\dot\al} \La_\al = 0,
\qquad
\bar{D}_{\dot\al} \La_{\be\dot\be} = -4i \varepsilon_{\dot\al \dot\be} \La_\be.
\eeq
The most general solution can be parameterized by
\beq[Lambdasoln]
\La_\al = -\sfrac 14 \bar{D}^2 L_\al,
\qquad
\La_{\al \dot\al} = -2i \bar{D}_{\dot\al} L_\al + \Om_{\al \dot\al},
\eeq
where $L_\al$ is an general complex superfield, and $\Om_{\al 
\dot\al}$ is chiral.
We will restrict attention to the transformations generated by $L_\al$.
(It can be checked that these form a closed subgroup.)
The transformation of a chiral superfield $\Phi$ can then be written
\beq
\de \Phi = -\sfrac 14 \bar{D}^2 (L^\al D_\al \Phi),
\qquad
\Phi = \hbox{\rm chiral}.
\eeq

The restricted group of super-diffeomorphisms we have found above does
not act naturally on antichiral superfields.
The constraints imposed by demanding that the general super-diffeomorphisms
in \Eq{superdiff} preserve antichiral fields are
\beq[firstconstr]
D_\al \La_{\dot\al} &= 0,
\\
\eql{secondconstr}
D_\al \La_{\be\dot\be} &= -4i \varepsilon_{\al\be} \La_{\dot\be}.
\eeq
\Eq{firstconstr} is satisfied by choosing
\beq[Lambdadot]
\La_{\dot\al} = (\La_\al)^\dagger = -\sfrac 14 D^2 \bar{L}_{\dot\al},
\eeq
but \Eq{secondconstr} is inconsistent with the chiral constraints
\Eq{chiralconstraint}, and cannot be imposed.
Following the construction of gauge theory, we define a covariantly
conjugate superfield%
%
%
\beq
\Phi^\ddagger \equiv (1 - 2i V^m \d_m) \Phi^\dagger
\eeq
such that $\Phi^\ddagger$
transforms according to the constrained super-diffeomorphisms
given by \Eqs{Lambdasoln} and \eq{Lambdadot}:
\beq
\de \Phi^\ddagger = -\sfrac 14 (D^2 \bar{L}_{\dot\al})
\bar{D}^{\dot\al} \Phi^\ddagger
+ i (\bar{D}^{\dot\al} L^\al) \partial_{\dot\al \al} \Phi^\ddagger.
\eeq
This requires
\beq
\de V_{\al\dot\al} = \bar{D}_{\dot\al} L_\al - D_\al \bar{L}_{\dot\al}.
\eeq
Note that $(\bar{D}_{\dot\al} L_\al)^\dagger = -D_\al \bar{L}_{\dot\al}$,
so $V_{\al\dot\al}$ is real.

To summarize, a general superfield $\Psi$ transforms according to 
\Eq{superdiff},
with $\La_\al$, $\La_{\dot\al}$, and $\La^m$ given by \Eqs{Lambdasoln}
and \eq{Lambdadot}.
It is sometimes convenient to make a field redefinition
\beq
\Psi' = (1 + 2i a V^m \d_m) \Psi,
\eeq
where $a$ is a real parameter.
The redefined field transforms as
\beq[gentrans]
\!\!\!\!\!\!\!
\de\Psi' = -\sfrac 14 (\bar{D}^2 L^\al) D_\al \Psi'
- \sfrac 14 (D^2 \bar{L}_{\dot\al}) \bar{D}^{\dot\al} \Psi'
+ i \bigl[ (1 - a) \bar{D}^{\dot\al} L^\al + a D^\al \bar{L}^{\dot\al} \bigr]
\d_{\al \dot\al} \Psi'.
\eeq
For $a = 0$, this preserves $\bar{D}_{\dot\al} \Psi' = 0$,
for $a = 1$ it preserves $D_\al \Psi' = 0$,
and for $a = \sfrac 12$ it preserves $\Psi'^\dagger = \Psi'$.
In this way we can define covariant versions of real, chiral, and
anti-chiral superfields.

\subsection{Components and the Chiral Compensator}
To better understand the gauge symmetries above, we consider their
action on the components of a chiral superfield $\Phi$.
We define the component fields by projection
\beq
\phi = \Phi |,
\qquad
\psi_\al = D_\al \Phi |,
\qquad
F = -\sfrac 14 D^2 \Phi |,
\eeq
where `$|$' denotes evaluation at $\th = 0$.
We then find
\beq[phitrans]
\de \phi &= \xi^m \d_m \phi + \varepsilon^\al \psi_\al,
\\
\eql{psitrans}
\de \psi_\al &= \eta^m_\al \partial_m \phi
+ \xi^m \d_m \psi_\al
+ \la_\al{}^\be \psi_\be
+ 2 \varepsilon_\al F,
\\
\de F &= \ka^m \partial_m \phi
+ \sfrac 12 \eta^m_\be \partial_m \psi^\be
+ \rho^\al \psi_\al
+ \xi^m \partial_m F
+ \la_\al{}^\al F,
\eeq
where
\beq
\xi^m &= i \tilde{\si}^{m\dot\al \al} \bar{D}_{\dot\al} L_\al |,
\\
\varepsilon_\al &= -\sfrac 14 \bar{D}^2 L_\al |,
\\
\eta^m_\al &= i \tilde{\si}^{m \dot\be \be} D_\al \bar{D}_{\dot\be} L_\be |,
\\
\la_\al{}^\be &= -\sfrac 14 D_\al \bar{D}^2 L^\be |,
\\
\ka^m &= -\sfrac i4 \si^m_{\al\dot\al} D^2 \bar{D}^{\dot\al} L^\al |,
\\
\rho_\al &= \sfrac{1}{16} D^2 \bar{D}^2 L_\al |.
\eeq
Note that the symmetrized generators $\la_{(\al\be)}$ generate local Lorentz
transformations.
The extra gauge symmetry in the trace
$\la_\al{}^\al$ generates scale and $U(1)_R$ transformations.
We see that gauging super-diffeomorphisms naturally gives rise to
{\it superconformal} supergravity.

%

To understand the components further, we go to Wess--Zumino gauge.
We define the components of the supergravity multiplet as
\beq
c_m &= V_m |,
\\
\chi_{\al\be\dot\be} &= D_\al V_{\be\dot\be} |,
\\
a_m &= -\sfrac 14 D^2 V_m |,
\\
h_{\al\dot\al, \be\dot\be} &= -\sfrac 12 [D_\al, \bar{D}_{\dot\al}]
V_{\be\dot\be} |,
\\
\eql{gravitinodefn}
\psi_\al^m &= 
\sfrac i{16} \tilde{\si}^{m\dot\be \be}
\bar{D}^2 D_\be V_{\al\dot\be} |,
\\
d_m &= \sfrac 1{32} \{ D^2, \bar{D}^2 \} V_m |.
\eeq
These transform as
\beq
\de c_m &= -\Im(\xi_m),
\\
\de \chi_{\al\be\dot\be} &= \sfrac i2 \si_{m \be\dot\be} \eta^m_\al
+ 2 \varepsilon_{\al\be} \bar{\varepsilon}_{\dot\be},
\\
\de a_m &= \sfrac i2 \ka_m,
\\
\de h_{\al\dot\al, \be \dot\be} &= \d_{\al\dot\al} \Re(\xi_{\be\dot\be})
- 2 (\varepsilon_{\dot\al \dot\be} \la_{\al\be}
+ \varepsilon_{\al\be} \bar{\la}_{\dot\al \dot\be}),
\\
\eql{gravitinotrans}
\de \psi^\al_m &= 
\d_m \varepsilon^\al
+ \sfrac i2 \tilde{\si}_m^{\dot\al \al}
\bar{\rho}_{\dot\al},
\\
\de d_m &= -\sfrac 12 \d^n \d_n \Im(\xi_m)
+ \bigl[\sfrac i4 (\tilde{\si}_m \si_n)_{\dot\al \dot\be} \d^n
\bar{\la}^{\dot\al \dot\be} + \hc \bigr].
\eeq
We use the gauge freedom in $\Im(\xi_m)$, $\eta^m_\al$,
and $\ka_m$ to set
\beq
c_m, \chi_{\al\be\dot\be}, a_m = 0.
\eeq
This leaves a residual gauge symmetry with
\beq
\eta^m_\al = 2i \si^m_{\al\dot\al} \bar{\varepsilon}^{\dot\al}.
\eeq
In this gauge, $\Re(\xi_m)$ generates local diffeomorphisms and
$\varepsilon$ generates local supersymmetry transformations on $\phi$ and
$\psi$.
The symmetric part of $\la_{\al\be}$ generates local Lorentz
transformations, as before.

We still have the `extra' gauge symmetry generated by $\la_\al{}^\al$ and
$\rho_\al$.
To fix this, we introduce the conformal compensator $\Si$.
This is a chiral field transforming as
%
%
\beq
\de \Si = -\sfrac 14 \bar{D}^2 D^\al L_\al,
\eeq
with components defined by
\beq
\si &= \Si |,
\\
\zeta_\al &= D_\al \Si |,
\\
F_\Si &= -\sfrac 14 D^2 \Si |.
\eeq
These components transform as
\beq
\de \si &= \d_m \xi^m - \la_\al{}^\al,
\\
\de \zeta_\al &= 2 \rho_\al +\d_m \eta^m_\al,
\\
\de F_\Si &= \d_m \ka^m.
\eeq
We can use the gauge symmetry generated by $\la_\al{}^\al$ and
$\rho_\al$ to set
\beq
\si = h^m{}_m ,
\qquad
\zeta_\al = 0.
\eeq
The first condition does not require any further compensating gauge 
transformation in Wess--Zumino gauge, where $\xi^m$ is real.
The second condition requires a compensating gauge transformation
\beq
\rho_\al = -\sfrac 12 \d_m \eta^m_\al = -i
\si^m_{\al\dot\al} \d_m \bar{\varepsilon}^{\dot\al},
\eeq
so that the gravitino transformation law \Eq{gravitinotrans} is modified
to
\beq
\de \psi^m_\al =
\left[ (\si^{mn})_\al{}^\be + \sfrac 32 \eta^{mn} \de_\al{}^\be \right]
\d_n \ep_\be.
\eeq
We can redefine the gravitino field to obtain a more conventional
transformation law (see \Eq{convgravitino} below).
However, working out the components in terms of the `unconventional'
gravitino field defined here is easier in this approach.

The remaining gauge symmetry in this gauge consists of diffeomorphisms
generated by $\Re(\xi^m)$,
local supersymmetry transformations generated by $\varepsilon_\al$,
and local Lorentz transformations generated by $\la_{(\al\be)}$.
These are precisely the gauge invariances we expect for supergravity.
The remaining supergravity component fields
\beq
h_{mn},\ \
\psi_m^\al,\ \
d_m,\ \
F_\Si
\eeq
comprise the minimal off-shell \N supergravity multiplet.

\subsection{Invariant Couplings}
We now construct the couplings of supergravity to chiral matter fields.
We first consider a superpotential term
\beq
\scr{L}_F = \myint d^2 \th\, W + \hc,
\eeq
where $W$ is a chiral superfield transforming as
$\de W = -\sfrac 14 \bar{D^2} (L^\al D_\al W)$.
Under super-diffeomorphisms we have
\beq
\de \! \myint d^2 \th W = \myint d^4\th\, L^\al D_\al W,
\eeq
so it is easy to see that adding
\beq
\De \scr{L}_F = \myint d^2\th\, \Si W + \hc
\eeq
makes the superpotential term invariant.

We next consider a \Kahler term
\beq
\scr{L}_D = \myint d^4\th\, K
\eeq
where $K$ is real and transforms like a real superfield
(\Eq{gentrans} with $a = \sfrac 12$).
For example, if $\Phi$ is a chiral superfield, then the minimal
kinetic term must be modified to
\beq
K = 
\Phi^\dagger \Phi + i V^m \Phi^\dagger \dd_m \Phi,
\eeq
where $A \dd B = A \d B - (\d A) B$.
Then
\beq
\de \! \myint d^4\th\, K = \myint d^4\th \bigl[
\sfrac 14 \bar{D}^2 D^\al L_\al + \sfrac i2 \d_{\al\dot\al} 
\bar{D}^{\dot\al} L^\al
+ \hc \bigr] \cdot K.
\eeq
This can be cancelled using the identity
\beq
\de \bigl(
[D_\al, \bar{D}_{\dot\al} ] V^{\dot\al \al} \bigr)
= -2 \bar{D}^2 D^\al L_\al
- 6i \d_{\al\dot\al} \bar{D}^{\dot\al} L^\al + \hc
\eeq
We therefore find that adding
\beq
\De \scr{L}_D
= \myint d^4\th \bigl\{ \sfrac 13 (\Si + \Si^\dagger)
  + \sfrac 1{12} [D_\al, \bar{D}_{\dot\al}] V^{\dot\al \al} \bigr\} \cdot K
\eeq
makes the \Kahler term invariant.


We now have enough results for 4D supergravity to tackle the 5D case.

\section{5D Supergravity}
In this section we construct minimal 5D supergravity in $\scr{N} = 1$
superspace.
On shell this theory contains a metric $g_{MN}$, a graviphoton $B_M$,
and a gravitino $\psi_M$, where $M = 0, \ldots, 3; 5$.
The on-shell theory was first constructed by Cremmer \cite{Cremmer}.
Since we will be interested only in the linear theory, the on-shell
action is simply the sum of the kinetic terms for the fields above.

\subsection{\label{embed}Superfield Embedding}
To find the superfields that parameterize the 5D supergravity fields,
we collect the bosonic fields and their gauge transformations.
The 5-bein fluctuation fields $h_{MN}$ transform as
\beq
\de h_{mn} &= \d_m \xi_n + \la_{mn},
\\
\de h_{5m} &= \d_5 \xi_m + \la_{5m},
\\
\de h_{m5} &= \d_m \xi_5 - \la_{5m},
\\
\de h_{55} &= \d_5 \xi_5,
\eeq
where we have used the fact that the generators of local Lorentz
transformations are antisymmetric $\la_{MN} = -\la_{NM}$.
The graviphoton fields transform as
\beq
\de B_m &= \d_m \al,
\\
\de B_5 &= \d_5 \al.
\eeq
The fields $h_{mn}$ are contained in the superfields
$V_m$ and $\Si$ of $\scr{N} = 1$ superfield supergravity, as reviewed in the previous
section.
We now discuss the superfield embedding of the remaining bosonic fields.

When this theory is compactified on an $S^1/Z_2$ orbifold, the zero modes
of $h_{55}$ and $B_5$ form a chiral `radion multiplet.'
It is therefore natural to parameterize these fields by a chiral superfield
\beq
T \sim h_{55} + i B_5 + \cdots = \hbox{\rm chiral}
\eeq
transforming as
\beq
\de T = \d_5 \Om,
\eeq
where
\beq
\Om \sim \xi_5 + i\al + \cdots = \hbox{\rm chiral}.
\eeq

The field $h_{m5}$ transforms as a gauge field with gauge parameter
$\xi_5$.
When 5D supergravity is compactified on $S^1$ this field parameterizes the
Kaluza--Klein gauge boson.
It is therefore natural to parameterize $h_{m5}$ by a real `Kaluza--Klein'
superfield
\beq
K \sim \th \si^m \bar{\th} h_{m5} + \cdots = \hbox{\rm real},
\eeq
transforming as
\beq
\de K = i(\Om - \Om^\dagger) - N,
\eeq
where
\beq
N \sim \th \si^m \bar{\th} \la_{5m} = \hbox{\rm real}.
\eeq
Note that the superfield transformation parameterized by $N$ can be
used to completely shift away $K$, just as the local Lorentz transformations
$\la_{5m}$ can be used to shift away $h_{m5}$.

We now turn to $h_{5m}$.
We do \emph{not} embed $h_{5m}$ in a real superfield
\beq
K' \sim \th \si^m \bar{\th} h_{5m} + \cdots = \hbox{\rm real},
\eeq
because $\xi_m$ is embedded in the superfield $L_\al$ as
\beq
L_\al \sim i\bar{\th}^{\dot\al} \xi_{\al\dot\al} + \cdots,
\eeq
and therefore the $\xi_m$ transformation law of $h_{5m}$ would
require
\beq
\de K' \sim i\th^\al \d_5 L_\al + \cdots.
\eeq
The appearance of explicit factors of superspace coordinates in
superfield transformations means that manifest global $\scr{N} = 1$
supersymmetry is lost.
We instead embed $h_{5m}$ and $B_m$ in a spinor superfield%
\footnote{The embedding of spin $\sfrac 32$ fields in superfields
of this type was first considered in \Ref{gates32}.}
\beq
\Psi_\al \sim \bar{\th}^{\dot\al} (B_{\al\dot\al} + i h_{5,\al\dot\al})
+ \cdots.
\eeq
We obtain the correct transformation law for $h_{5m}$ and $B_m$ if we take
$\Psi_\al$ to transform as
\beq
\de \Psi_\al = \d_5 L_\al + \sfrac{i}{4} D_\al N.
\eeq

We now collect the 5D supergravity $\scr{N} = 1$
superfields and their complete
transformation laws:
\beq
\de T &= \d_5 \Om,
\\
\de K &= i (\Om - \Om^\dagger) - N,
\\
\de \Psi_\al &= \d_5 L_\al + \sfrac i4  D_\al N.
\eeq
As already noted above, we can use the $N$ gauge transformation to completely
shift away $K$.
Equivalently, the action depends on $K$ only through the combination
\beq
\hat\Psi_\al = \Psi_\al + \sfrac i4 D_\al K,
\eeq
which transforms as
\beq
\de\hat{\Psi}_\al = \d_5 L_\al - \sfrac 14 D_\al \Om.
\eeq





\subsection{Invariant Action}

We now write an invariant lagrangian for the supergravity fields
found above.
This can be constructed systematically by working
order by order in $\d_5$.
We find that in order to write an invariant action we must introduce
a prepotential $P$ for the conformal compensator.
(This prepotential for 4D supergravity was previously introduced in
\Ref{gatesP}.)
We take $P$ to be real and write
\beq
\Si = -\sfrac 14 \bar{D}^2 P,
\qquad
\de P = D^\al L_\al + \hc 
\eeq
We then find that the most general invariant lagrangian that is
quadratic in the 5D supergravity fields is
\beq[SUGRA5D]
\scr{L} = \scr{L}_{\scr{N} = 1}
+ c \De{\scr{L}}_5,
\eeq
where $\scr{L}_{\scr{N} = 1}$ is the lagrangian of linearized $\scr{N} = 1$
supergravity \cite{buch}:
\beq[LN1]
\bal
\scr{L}_{{\cal N} = 1} &=
\myint d^4\th \Bigr[
\sfrac 18 V^m D^\al \bar{D}^2 D_\al V_m
+ \sfrac 1{48} \left( [D^{\al}, \bar{D}^{\dot\al} ] V_{\al\dot\al} \right)^2
- (\d^m V_m)^2
\\
& \qquad\qquad
- \sfrac 13 \Si^\dagger \Si
+ \sfrac {2i}3 (\Si - \Si^\dagger) \d^m V_m \Bigr].
\eal\eeq
and
\beq[L5]
\bal
\De{\scr{L}}_5 = \myint d^4\th \Bigl\{ &
\bigl[  T^\dagger ( \Si - i \d_{\al \dot\al} V^{\dot\al \al}) + \hc \bigr]
\\
& - \sfrac 12 \bigl[ D^\al \hat\Psi_\al
+ \bar{D}_{\dot\al} \hat{\Psi}^{\dagger \dot\al}
- \d_5 P \bigr]^2
\\
& + \bigl[ \d_5 V_{\al\dot\al} - (\bar{D}_{\dot\al} \hat\Psi_\al
- D_\al \hat\Psi^\dagger_{\dot\al} ) \bigr]^2 \Bigr\},
\eal
\eeq
The constant $c$ is to be determined by imposing 5D Lorentz invariance.

If we write the action as $S = M_5^3 \int d^5 x \scr{L}$, then the 
superfields have
the following mass dimensions:
\beq{}
[V] = -1,
\quad
[P] = -1,
\quad
[\hat{\Psi}] = -\sfrac 12,
\quad
[T] = 0.
\eeq
With this convention, propagating bosonic fields have dimension $0$.

\subsection{Components}
We now work out the bosonic part of the lagrangian \Eq{SUGRA5D}.
We define the components by covariant projection, as before.
Our definitions are chosen to obtain simple transformation laws.

The bosonic components of $T$ are defined to be
\beq
t &= T|,
\\
F_T &= -\sfrac 14 D^2 T |.
\eeq
The bosonic components of $\hat{\Psi}_\al$ are defined to be
\beq
u_{\al\be} &= D_\al \hat{\Psi}_\be |,
\\
v_{\al\dot\al} &= -2i \bar{D}_{\dot\al} \hat{\Psi}_\al |,
\\
w_{\al\be} &= -\sfrac 14 D_\al \bar{D}^2 \hat{\Psi}_\be |,
\\
y_{\al\dot\al} &= -\sfrac 14 D^2 \bar{D}_{\dot\al} \hat{\Psi}_\al |.
\eeq
The bosonic components of $V_{\al\dot\al}$ are defined as before:
\beq
c_{\al\dot\al} &= V_{\al\dot\al} |,
\\
a_{\al\dot\al} &= -\sfrac 14 D^2 V_{\al\dot\al} |,
\\
\tilde h_{\al\dot\al \be\dot\be} &= -\sfrac 12 [D_\al, \bar{D}_{\dot\al}]
V_{\be\dot\be} |,
\\
d_{\al\dot\al} &= \sfrac 1{32} \{ D^2, \bar{D}^2 \} V_{\al\dot\al} |.
\eeq
The bosonic components of the prepotential $P$ are
\beq
\varrho &= P |,
\\
\si &= -\sfrac 14 \bar{D}^2 P |,
\\
\tau_{\al\dot\al} &= -\sfrac 12 [D_\al, \bar{D}_{\dot\al} ] P |,
\\
D_P &= \sfrac 1{32} \{ D^2, \bar{D}^2 \} P|.
\eeq
The relationship between $D_P$ and $F_\Si$ is
\beq
F_\Si = D_P - \sfrac i2 \d_m \tau^m.
\eeq

We first work out the transformation of these components under the gauge
symmetries.
We define the bosonic components of the transformation parameter $\Om$
as
\beq
\om &= \Om |,
\\
F_\Om &= -\sfrac 14 D^2 \Om |.
\eeq
The bosonic components of $L_\al$ are
\beq
\ga_{\al\be} &= D_\al L_\be |,
\\
\xi_{\al\dot\al} &= -2i \bar{D}_{\dot\al} L_\al |,
\\
\la_{\al\be} &= -\sfrac 14 D_\al \bar{D}^2 L_\be |,
\\
\ka_{\al\dot\al} &= \sfrac i2 D^2 \bar{D}_{\dot\al} L_\al |.
\eeq
The bosonic components of $T$ transform as
\beq
\de t &= \d_5 \om,
\\
\de F_T &= \d_5 F_\Om.
\eeq
The bosonic components of $\Psi_\al$ transform as
\beq
\de u_{\al\be} &= \d_5 \ga_{\al\be} + \sfrac 12 \ep_{\al\be} F_\Om,
\\
\de v_m &= \d_5 \xi_m + \d_m \om,
\\
\de w_{\al\be} &= \d_5 \la_{\al\be},
\\
\de y_m &= \sfrac i2 \d_5 \ka_m + \sfrac i2 \d_m F_\Om.
\eeq
The bosonic components of $V_{\al\dot\al}$ transform as
\beq
\de c_m &= -\Im(\xi_m),
\\
\de a_m &= \sfrac i2 \ka_m,
\\
\de \tilde h_{\al\dot\al \be \dot\be} &= \d_{\al\dot\al} \Re(\xi_{\be\dot\be})
- 2 (\varepsilon_{\dot\al \dot\be} \la_{\al\be}
+ \varepsilon_{\al\be} \bar{\la}_{\dot\al \dot\be}),
\\
\de d_m &= -\sfrac 12 \d^n \d_n \Im(\xi_m)
+ \bigl[\sfrac i4 (\tilde{\si}_m \si_n)_{\dot\al \dot\be} \d^n
\bar{\la}^{\dot\al \dot\be} + \hc \bigr].
\eeq
The bosonic components of $P$ transform as
\beq
\de \varrho &= \ga^\al{}_\al + {\rm h.c.},
\\
\de \tau_{\al\dot\al} &= -2 \Im(\ka_{\al\dot\al}) 
+ \bigl[ i \d^\be{}_{\dot\al} (\ga_{\al\be} + \ga_{\be\al}) + \hc \bigr],
\\
\de \si &= \d_m \xi^m - \la_\al{}^\al,
\\
\de D_P &= \d^m \Re(\ka_m).
\eeq

We now work out the bosonic terms in the lagrangian to check 5D global
Lorentz invariance and determine the constant $c$ in \Eq{SUGRA5D}.
We use the gauge freedom in
$\ga^\al{}_\al$, $\la^\al{}_\al$, $\Im(\xi_m)$, $\ka_m$, and $F_\Om$ to
go to a Wess--Zumino gauge where
\beq
\varrho, \si, c_m, a_m, u^\al{}_\al = 0.
\eeq
This gauge eliminates all bosonic fields with dimension less than $0$,
the dimension of propagating bosonic fields in our conventions.
The gauge condition $\si = 0$ requires a compensating gauge transformation,
so that $\tilde h_{mn}$ does not transform like a canonical
5-bein fluctuation in this gauge.
We will make contact with more familiar expressions by writing our
final results in terms of the canonical field $h_{mn}$, given by
\beq
\tilde h_{(mn)} = h_{mn} - \eta_{mn} h^p{}_p.
\eeq

The component lagrangian in this gauge is
\beq\bal
\scr{L}_{\scr{N} = 1} &= -\sfrac 12 (\d_m \tilde h_{np})^2
+ \sfrac 12 (\d^m \tilde h_{mn})^2
+ \sfrac 12 (\d^m \tilde h_{nm})^2
\\
& \qquad
- \sfrac 16 (\Om_m)^2
+ \sfrac 16 (\d_m \tilde h)^2
+ \sfrac 13 \tilde h \d_m \d_n \tilde h^{mn}
\\
& \qquad
+ \sfrac 23 d^m \Om_m
- \sfrac 13 |F_\Si|^2
+ \sfrac 43 (d_m)^2,
\eal\eeq
and
\beq\bal
\!\!\!\!\!
\De\scr{L}_5 &= 2 \Re(F_T) D_P
- \Im(F_T) \d_m \tau^m
- 4 d^m \d_m \Im(t)
- 2 \Re(t) \d_m \d_n \tilde h^{mn}
\\
& \qquad
- \left| \d_m v^m + w^\al{}_\al \right|^2
- 2 \left[ \Re(y_{\al\dot\al})
- \left( \sfrac i2 \d_{\be\dot\al} u^\be{}_\al + \hc \right)
- \sfrac 14 \d_5 \tau_{\al\dot\al} \right]^2
\\
& \qquad
- \Im(v^m) \d^2 \Im(v_m)
- 4 d^m \d_5 \Im(v_m)
- 4 | y_m |^2
\\
& \qquad
+ \sfrac 14 \left[ \d_{\al\dot\al} \Re(v_{\be\dot\be})
- \d_5 \tilde h_{\al\dot\al,\be\dot\be} - 4 
\Re(\varepsilon_{\dot\al\dot\be}) \right]^2
\\
& \qquad
- \left( i w_{\be\al} \d^\be{}_{\dot\al} \Im(v^{\dot\al \al}) + \hc \right),
\eal\eeq
where $\tilde h = \tilde h^m{}_m$ and
\beq
\Om_m = \varepsilon_{mnpq} \d^n \tilde h^{pq}.
\eeq

We now discuss the elimination of the auxiliary fields.
The equations of motion of the auxiliary fields $F_T$ and $D_P$ set
\beq
F_T, D_P, \d^m \tau_m = 0.
\eeq
The equations of motion of the auxiliary field $y_m$ (from $\hat{\Psi}_\al$)
eliminate all
dependence on the fields $\tau_m$ and $u_{\al\be}$:
\beq\bal
\scr{L} &= c \Bigl\{ \Re(y^{\dot\al \al}) \left[
(2i \d_{\be \dot\al} u^\be{}_\al + \hc) + \d_5 \tau_{\al\dot\al} \right]
+ \sfrac 12 (\Im(y^m))^2  \Bigr\}
\\
& \qquad\quad
+ \hbox{\rm independent\ of\ }y.
\eal\eeq
This is important because $\tau_m$ and $u_{\al\be}$ have dimension $0$,
the dimension of a propagating bosonic field.
The remaining fields of dimension $0$ are just enough to parameterize the
bosonic component fields of the theory (see \S\ref{embed} above).

The only remaining auxiliary fields are $d_m$ and $w_{\al\be}$.
Their equations of motion give
\beq
d_m &= \sfrac 32 c \left[ \d_m \Im(t) - \d_5 \Im(v_m) \right]
- \sfrac 14 \Om_m,
\\
w_{\al\be} &= -\sfrac 13 \ep_{\al\be} \d_5 \tilde h
+ \sfrac 18 \left[ \d_{(\al}{}^{\dot\al} \Re(v_{\be)\dot\al})
- \d_5 \tilde h_{(\al}{}^{\dot\al}{}_{\be)\dot\al}
- i \d_{(\al}{}^{\dot\al} \Im(v_{\be)\dot\al}) \right].
\eeq
Substituting these back into the action results in a large number of
cancellations.
In particular, terms of the form
$\d_m \Re(v^m)\d_5 \tilde h$,
$\Om^m \d_5 \Im(v_m)$,
and $(\Om_m)^2$ cancel. Additionally, all further dependence on the 
antisymmetric part of the 5-bein fluctuation $\tilde h_{[mn]}$ 
cancels and we are left with
\beq\bal
\scr{L} &= -\sfrac 12 (\d_m \tilde h_{(np)})^2
+ (\d^m \tilde h_{(mn)})^2
+\sfrac 16 (\d_m \tilde h)^2
+ \sfrac 13 \tilde h \d_m \d_n \tilde h^{(mn)}
\\
& \qquad - \sfrac 13 c (\d_5 \tilde h)^2
+ c (\d_5 \tilde h_{(mn)})^2
- 2 c \Re(t) \d_m \d_n \tilde h^{(mn)}
\\
& \qquad
+ \sfrac 14 c \left[ \d_m \Re(v_n) - \d_n \Re(v_m) \right]^2
- 2 c \d_m \Re(v_n) \d_5 \tilde h^{(mn)}
\\
& \qquad
+ \sfrac 34 c \left[ \d_m \Im(v_n) - \d_n \Im(v_m) \right]^2
- 3 c^2 \left[ \d_m \Im(t) - \d_5 \Im(v_m) \right]^2,
\eal\eeq
The last two terms give the graviphoton gauge kinetic
term.
5-dimensional Lorentz invariance of these terms requires
\beq
c = -\sfrac 12.
\eeq
The lagrangian is now completely fixed, and gives a 5D Lorentz invariant
result.
It is convenient to express the final result in terms of the fields
\beq
h_{mn}&=\tilde h_{(mn)}-{\sfrac 13}\eta_{mn} \tilde h^p{}_p,
\\
h_{m5} &= \sfrac 12 \Re(v_m),
\\
h_{55} &= \Re(t),
\\
B_m &= \sqrt{\sfrac 32} \Im(v_m),
\\
B_5 &= \sqrt{\sfrac 32} \Im(t),
\eeq
with transformation laws
\beq
\de h_{mn} &= \sfrac 12 \left( \d_m \xi_n + \d_n \xi_m \right),
\\
\de h_{m5} &= \sfrac 12\left( \d_m \xi_5 + \d_5 \xi_m\right),
\\
\de h_{55} &= \d_5 \xi_5,
\\
\de B_m &= \d_m \al,
\\
\de B_5 &= \d_5 \al,
\eeq
where
\beq
\xi_5 &={\rm Re}(\omega),
\\
\al &= \sqrt{\sfrac 32} {\rm Im}(\omega).
\eeq
In terms of these fields 
the lagrangian is
\beq\bal
\scr{L} &= -\sfrac 12 (\d_m h_{np})^2
+ (\d^m h_{mn})^2
+\sfrac 12 (\d_m h)^2
+  h \d_m \d_n h^{mn}
\\
& \qquad + \sfrac 12  (\d_5 h)^2
-\sfrac 12 (\d_5 h_{mn})^2
+ h_{55} \d_m \d_n h^{mn}
\\
& \qquad
- \sfrac 12  \left[ \d_m h_{n5} - \d_n h_{m5} \right]^2
+ 2  \d_m h_{n5} \d_5 h^{mn}-2 \d^m h_{m5} \d_5 h
\\
& \qquad
- \sfrac 14  \left[ \d_m B_n - \d_n B_m \right]^2
-\sfrac 12 \left[ \d_m B_5 - \d_5 B_m \right]^2.
\eal
\eeq
This is the correct 5D lagrangian for linearized gravity plus a
graviphoton field.

\subsection{Fermions and 5D Supersymmetry}
We will not carry out the component expansion of the fermions,
but we will show that the theory contains all components of the
5D gravitino with the correct transformation law under
infinitesimal local 5D supersymmetry.
In the 5D theory, the gravitino $\Psi_M$ transforms under local
supersymmetry as
\beq
\de \Psi_M = \d_M \varepsilon.
\eeq
In the reduction to $\scr{N} = 1$ superspace, the gravitino decomposes into the
fields
$\psi^{(\pm)}_{M \al}$,
where the $\pm$ refers to the intrinsic parity under $x^5 \mapsto -x^5$.

Propagating fermion fields have mass dimension $+\sfrac 12$ in our
conventions.
The propagating fermion fields are
\beq
\eql{convgravitino}
\psi_{m\al}^{(+)} &=
\left[ -\sfrac 13 (\si_{mn})_\al{}^\be
+ \sfrac 56 \eta_{mn} \de_\al{}^\be \right]
\psi^n_\be,
\\
\psi_{m\al}^{(-)} &=i(\tilde \sigma_m)^{\dot \beta \beta}D_\alpha \bar 
D_{\dot \beta} \hat \Psi_\beta |- \sfrac i2 (\sigma_m)_\alpha{}^{\dot 
\beta}D^2{\hat {\bar \Psi}}_{\dot \beta}|.
\\
\psi_{5\al}^{(+)} &=-\sfrac 14 \bar D^2 \hat \Psi_\alpha |.
\\
\psi_{5\al}^{(-)} &= D_\alpha T|,
\eeq
where $\psi_{m\al}$
is the `unconventional' gravitino field defined in \Eq{gravitinodefn}.
These transform as
\beq
\de \psi_{M \al}^{(\pm)} = \d_M \varepsilon^{(\pm)}_\al,
\qquad M = 0, \ldots, 3, 5,
\eeq
where the transformation parameters are
\beq
\varepsilon_{\al}^{(+)} &=-\sfrac 14 \bar D^2 L_\alpha|,
\\
\varepsilon_{\al}^{(-)} &=D_\alpha \Omega |.
\eeq

\section{Applications}

\subsection{$S^1 / Z_2$ Orbifold}
We now consider the compactification of this theory on an $S^1/Z_2$
orbifold.
This is a good starting point for constructing realistic `brane world'
scenarios, since this compactification breaks supersymmetry down to
$\scr{N} = 1$ and gives rise to two `branes' at the orbifold fixed points.
In the present formalism, the $Z_2$ parity assignments are simply
\beq
\scr{P}(V_m) = +1,
\qquad
\scr{P}(P) = +1,
\qquad
\scr{P}(\Psi_\al) = -1,
\qquad
\scr{P}(T) = +1.
\eeq
It is now simple to couple the 5D supergravity multiplet to matter fields
localized on the boundaries, since the
5D supergravity multiplet induces a $\scr{N} = 1$ minimal supergravity multiplet
on the boundaries at $x^5 = 0, \pi r$.
Note that the radion superfield transforms as
\beq
\de T = \d_5 \Om,
\eeq
and $\d_5 \Om$ is a general chiral superfield on the boundary.
Since superfields localized on the boundary do not transform under $\Om$,
we see that there are no couplings of the radion superfields to the boundary.
(This result was obtained in \Ref{LS} by an indirect argument.)

Since we have derived the linearized theory, our results apply
directly to supergravity couplings to the boundary at linear order.
However, it is clear that the fully nonlinear form of these
couplings is simply obtained by including the standard nonlinearization
of the induced $\scr{N} = 1$ supergravity multiplet.
Of particular importance is the conformal compensator, given by
\beq
\phi = e^{\Si / 3}.
\eeq
We can now use the usual nonlinear couplings of $\scr{N} = 1$ supergravity
to 4D matter.
For example, the couplings of brane localized fields to the conformal
compensator are given by
\beq
\de \scr{L}_5 = \de(x^5) \left[
\myint d^4\th\, \phi^\dagger \phi f
+ \left( \myint d^2\th\, \phi^3 W + \hc \right) \right]
+ \cdots
\eeq
where $f$ is the \Kahler function and $W$ is the superpotential,
and we omit the dependence on $V_m$.

\subsection{The $\mu$ Term from 5D Supergravity}
As an application of this formalism, we show that couplings of
5D supergravity to branes
can naturally generate a $\mu$ term of realistic size in the context of
gaugino mediation \cite{gMSB}.
This can be viewed as a 5D version of the Giudice--Masiero
mechanism \cite{gm}.

We consider a theory with standard model Higgs fields $H_{u,d}$
localized on the boundary at $x^5 = 0$.
Consider the following brane-localized term added to the 5-dimensional
lagrangian:
\beq
\De \scr{L}_5 = \de(x^5) \myint d^4\th\, \phi^\dagger \phi
\left[ H_u^\dagger H_u + H_d^\dagger H_d
+ (c H_u H_d + \hc) + \cdots \right].
\eeq
where $c$ is a dimensionless coupling.
We have omitted the dependence on the supergravity field $V_m$ but we have
given the full dependence on the conformal compensator $\phi$.
Supersymmetry breaking gives rise to $\avg{F_\phi} \ne 0$ and
generates effective $\mu$ and $B \mu$ terms
\beq
\mu = c \avg{F_\phi^\dagger},
\qquad
B \mu = -c | \avg{F_\phi} |^2.
\eeq
In gaugino mediation, supersymmetry breaking is communicated to the
bulk gaugino via boundary couplings of the form
\beq
\De \scr{L}_5 = \de(x^5 - \pi r) \myint d^2\th\,
\frac{X}{M} W^\al W_\al + \hc + \cdots,
\eeq
where $X$ is a chiral superfield whose $F$ component breaks supersymmetry,
and $W_\al$ is the field strength of the bulk gauge multiplet.
To estimate the size of $M$ we assume that at the cutoff of the theory
$\La$ where all loop effects are suppressed by a factor of
$\ep \lsim 1$ \cite{NDA}.
This gives
\beq[gmsbest1]
\La \sim M_4,
\qquad
2\pi r \sim \frac{24 \pi^3 \ep}{M_4},
\eeq
and
\beq[gmsbest2]
c \sim 1,
\qquad
M \sim 4\pi \sqrt\ep M_4.
\eeq
This implies that $B \sim \mu$, with
\beq
\frac{\mu}{m_{1/2}} \sim \frac{1}{4\pi \sqrt{\ep}}.
\eeq
This gives $\mu$ and $B\mu$ terms of the right order of
magnitude for $10^{-2} \lsim \ep \lsim 1$.
Sequestering requires
$2 \pi r \La \sim 24\pi^3 \ep \gsim 7$, which is satisfied for all
$\ep$ in this range.

Similar results hold for radion-mediated supersymmetry breaking
(third paper in \Ref{gravmed}).
\Eqs{gmsbest1} and \eq{gmsbest2} are still valid, but
$m_{3/2} \sim \avg{F_T / T} \sim \sfrac 1{10}$.
This gives $B \sim \mu$ with
\beq
\frac{\mu}{m_{1/2}} \sim 10.
\eeq
This may be acceptable given the uncertainties in the estimates above.

\section{Conclusions}
To summarize,
we have presented an embedding of linearized 5D supergravity
into $\scr{N} = 1$ superfields, that is functions of 4D superspace
$(x^m, \th_\al)$, together with $x^5$.
The propagating fields are embedded in superfields as follows.
The fields $h_{mn}$ and $\psi^{(+)}_{m\al}$ are embedded in the
standard $\scr{N} = 1$ supergravity superfield
\beq
V_m \sim \th \si^n \bar\th h_{mn} + \bar{\th}^2 \psi^{(+)}_{m \al} 
+ \cdots = \hbox{\rm real},
\eeq
and the remaining propagating fields are embedded into the superfields
\beq
\Psi_\al &\sim \bar{\th}^{\dot\al} \si^m_{\al\dot\al} (B_m + i h_{5m})
+ \th \si^m \bar\th \psi^{(-)}_{m \al}
+ \bar{\th}^2 \psi_{5 \al}^{(+)} + \cdots
= \hbox{\rm unconstrained},
\\
T &\sim h_{55} + i B_5 + \th^\al \psi_{5 \al}^{(-)} + \cdots
= \hbox{\rm chiral},
\eeq
in a gauge where $h_{5m} = h_{m5}$.
Additionally, the theory contains a real superfield $P$ that acts as
a prepotential for the conformal compensator.
On an $S^1/Z_2$ orbifold, $V_m$, $P$, and $T$ are even, while
$\Psi_\al$ is odd.
The 5D lagrangian is given in \Eqs{LN1} and \eq{L5}.
The induced supergravity multiplet on the boundary is the usual
$\scr{N} = 1$ supergravity multiplet, so coupling to boundary
supermultiplets is simple.

We believe that this formalism will be useful in systematically
including supergravity effects in higher-dimensional theories
and `brane-world' scenarios.
As a first step in this direction, we have shown how couplings of 5D
supergravity to boundary Higgs fields can give realistic $\mu$
and $B \mu$ terms in the context of gaugino- and radion-mediated
supersymmetry breaking.
There are numerous open
directions for future work.
These include the generalization to other
backgrounds (such as `warped' compactifications), the extension beyond linear
order, coupling to bulk hypermultiplets and gauge multiplets,
and generalizations to dimensions higher than 5.

\section*{Acknowledgements}
We thank S. James Gates, Jr.~and R.~Sundrum for useful discussions.
W.D.L. and J.P. were supported by the
University of Maryland Center
for String and Particle Theory.
M.A.L. was supported by NSF grant PHY-0099544.

\end{document}